# Comparison of Maximum Likelihood Classification Before and After Applying Weierstrass Transform

Muhammad Shoaib, Zaka Ur Rehman, Muhammad Qasim

***Abstract*— The aim of this paper is to use Maximum Likelihood (ML) Classification on multispectral data by means of qualitative and quantitative approaches. Maximum Likelihood is a supervised classification algorithm which is based on the Classical Bayes theorem. It makes use of a discriminant function to assign pixel to the class with the highest likelihood. Class means vector and covariance matrix are the key inputs to the function and can be estimated from training pixels of a particular class. As Maximum Likelihood need some assumptions before it has to be applied on the data. In this paper we will compare the results of Maximum Likelihood Classification (ML) before apply the Weierstrass Transform and apply Weierstrass Transform and will see the difference between the accuracy on training pixels of high resolution Quickbird satellite image. Principle Component analysis (PCA) is also used for dimension reduction and also used to check the variation in bands. The results shows that the separation between mean of the classes in the decision space is to be the main factor that leads to the high classification accuracy of Maximum Likelihood (ML) after using Weierstrass Transform than without using it.**

## I. Introduction

Satellite images are widely used for various purposes and with the advancement in Satellite technology every country wants to solve their problem by using satellite images. Satellite images can be used in various disciplines such as Metrology, Environmental sciences, GIS, Geology, Geophysics, Engineering and Construction, Defense and Intelligence and the list goes on.

Satellite images contains maximum information about our planet and main problem is that how can we retrieve the useful information satellite images. Researchers can have several problems such how to classify the land cover types, depth of Aerosol particles in atmosphere, Forest Climates, Flood predictions and many more. Researcher solve these problems by analyzing the satellite images.

In this article our main aim is to classify the land cover types by using Maximum Likelihood Classification algorithm. First we classify land cover types without using Weierstrass Transform (WT) and then we compare the results by applying Weierstrass Transform (WT).

Maximum Likelihood (ML) is a supervised classification method derived from the Bayes theorem, which states that the a posteriori distribution $P(i|\theta)$, i.e., the probability that a pixel with feature vector $\theta$ belongs to class $i$, is given by:

$$p(i|\theta) = \frac{p(\theta|i) * p(i)}{p(\theta)} \quad \rightarrow \quad (1)$$

where $P(\theta|i)$, is the likelihood function, $P(i)$ is the a priori information, i.e., the probability that class $i$ occurs in the study area and $P(\theta)$ is the probability that is observed, which can be written as:

$$P(\theta) = \sum_{i=1}^{M} P(i|\theta)p(i) \quad \rightarrow \quad (2)$$

where $M$ is the number of classes. $P(\theta)$ is often treated as a normalization constant to ensure $\sum P(i|\theta)$ sums to 1. Pixel $x$ is assigned to class $i$ by the rule:

$$x \in i \quad if\, P(i|\theta) > P(j|\theta) \; for\, all\, i \neq j \quad \rightarrow \quad (3)$$

Each pixel is assigned to the class with the highest likelihood or labeled as unclassified if the probability values are all below a threshold set by the user [1].

Maximum Likelihood (ML) assumes that the data within a given class $i$ obeys a multivariate Gaussian distribution.

Frist we implement Maximum Likelihood (ML) classification algorithm without checking the assumption of normality and then we apply Maximum Likelihood (ML) classification after using Weierstrass Transform on the image data and we have seen that after applying Weierstrass Transform our data is normally distributed. The accuracy after applying Weierstrass Transform (WT) is increased because the distribution of the data is normal because for ML classification data has to be normal.

## II. Study Area

We use Quickbird satellite image for our analysis. Quickbird is high-resolution commercial satellite which is owned by DigitalGlobe and launched in 2001. Quickbird uses Aerospace's Global Imaging System 2000 (BGIS 2000).The Quickbird satellite collected panchromatic (black and white) at 61cm resolutin and multispectral imagery at 2.44-(at 450 km) to 1.63-meter (at 300km) resolution, as orbit altitude is lowered during the end of mission life [2].

The Quickbird satellite image which we use for our analysis is sample image of Rome Italy with coordinates $12^{o}34'21.43''\, E, 41^{o}52'25.49''\, N$ with four bands which are blue (450-520nm), green (520-600nm), red (630-690 nm) and NIR (760-900 nm).

## III. Methodology

Methodology is the systematic, theoretical analysis of the methods applied to a field of study. It comprises the theoretical analysis of the body of methods and principles associated with a branch of knowledge (wiki).

The detail methodology is given in the figure 1.

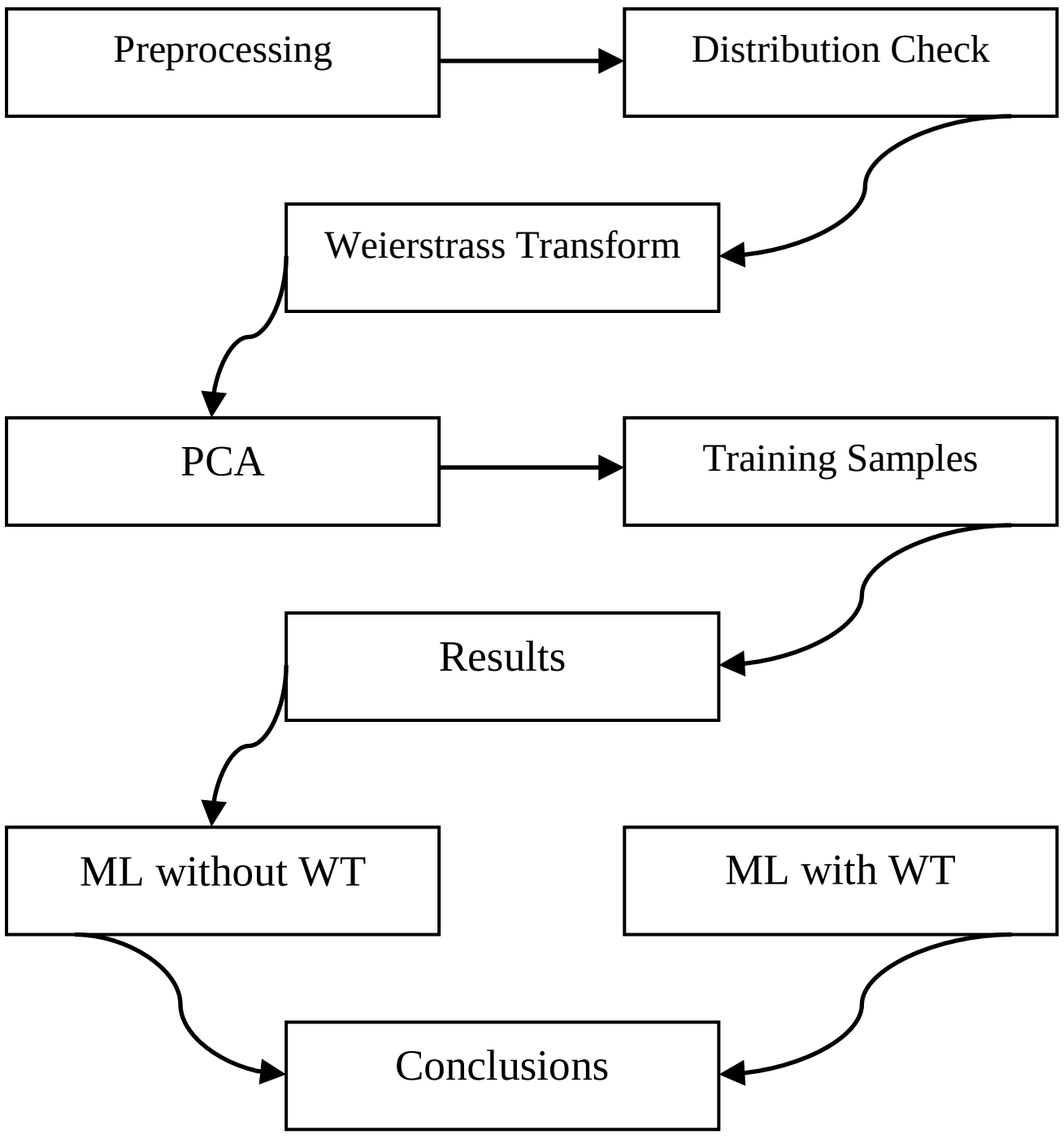

Fig. 1. Data Processing flowchart

## IV. Preprocessing

Preprocessing is a common name for operations with images at lowest level of abstraction both input and output are intensity images. The main aim of preprocessing is an improvement of the image data that suppresses unwanted distortions or enhances some image features important for further processing.

In this paper we have to do classification so we do not need to do any radiometric calibration or anything else. But we have to check the distribution of each band whether it is normal or not which is the main assumption for ML classification.

### *A. Distribution Check*

We use image histogram to check the distribution of each band. An "image histogram" is a type of histogram that acts as a graphical representation of the tonal distribution in a digital image. It plots the number of pixels for each tonal value. By looking at the histogram for a specific image a viewer will be able to judge the entire tonal distribution at a glance. We make a separate histogram for each band to check the distribution of each band in image.

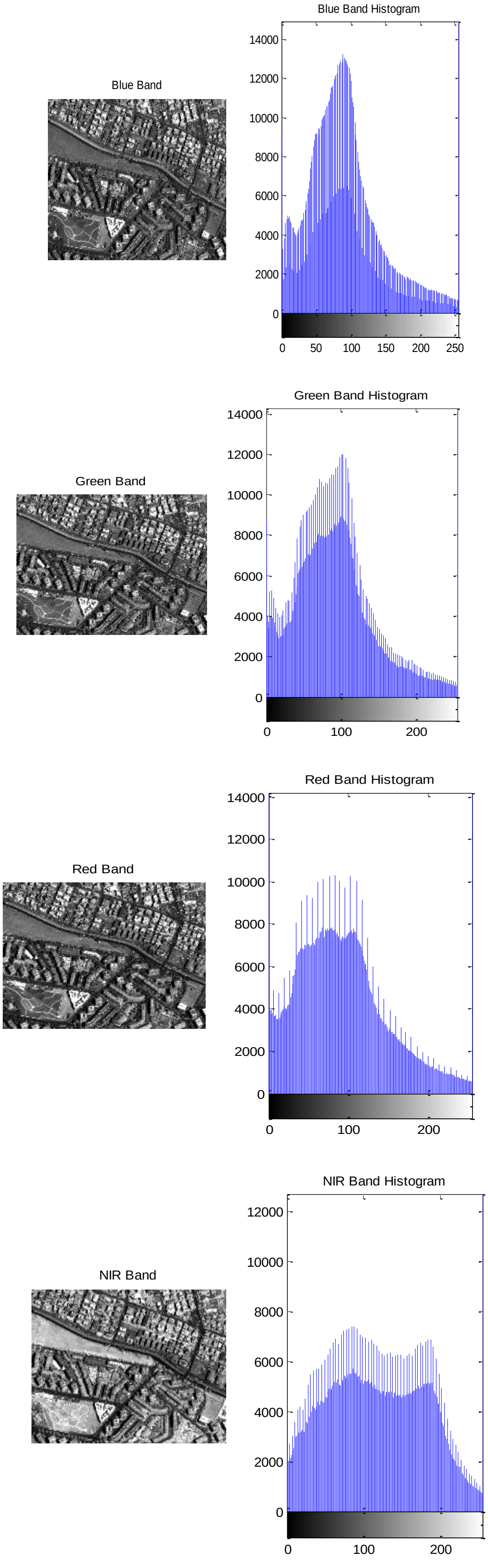

Fig. 2. Histograms with their respective bands

In the figure 2 we can see that the distribution of each band is not Gaussian and we know that for ML classification algorithm the distribution of each has to be Gaussian. In this case distribution of first three bands are positively skewed while the histogram of band 4 is looks like Gaussian but not exactly Gaussian. So as our aim is to check the accuracy of classification before making data Gaussian and after making it Gaussian. We use Weierstrass Transform to make the distribution Gaussian.

*B. Weirstrass Transform*

As we have seen that distribution of our data is not follow Gaussian distribution so make that Gaussian we have to use transformation to make pixel values Gaussian.

Researchers uses various types of transformation to make the data Gaussian. In this paper we use different type of transformation called Weirstrass Transform.

The Weirstrass transformation, also known as the Gauss transform, the Gauss Weirstrass transform and the Hille transform, is intimately related to the solution of the heat equation for one- dimensional flow. It is also a special case of convolution-transform, yet it is considered a singular case of convolution-transform theory developed by Hirschmann and Widder [3].

The Weirstrass transformation ***F(x)*** of a function ***f(y)*** defined as:

$$F(x) = \frac{1}{\sqrt{4\pi}} \int_{-\infty}^{\infty} f(y) e^{-\frac{(x-y)^2}{4}} dy$$

$$= \frac{1}{\sqrt{4\pi}} \int_{-\infty}^{\infty} f(y) e^{-\frac{y^2}{4}} dy$$

the convolution of ***f*** with the Gaussian function $\frac{1}{\sqrt{4\pi}} e^{-\frac{x^2}{4}}$. Instead of ***F(x)*** we also write ***W[f](x)***. Note that ***F(x)*** need not exist for every real number ***x***, because the defining integral may fail to converge ***[3]***.

When Gaussian filter modifies the input signal by using Gaussian function; this is known as Weierstrass transformation.

We can write Gaussian function mathematically in 2D as follow:

$$f(x,y) = \frac{1}{2\pi\sigma_x\sigma_y\sqrt{1-\rho^2}} \exp\left(-\frac{1}{2(1-\rho^2)}\left[\frac{(x-\mu_x)^2}{\sigma_x^2} + \frac{(y-\mu_y)^2}{\sigma_y^2} - \frac{2\rho(x-\mu_x)(y-\mu_y)}{\sigma_x\sigma_y}\right]\right) \rightarrow (1)$$

where ***ρ*** is the correlation between ***X*** and ***Y*** and where ***σ_x>0*** and ***σ_y>0***. In this case,

$$\mu = \begin{pmatrix} \sigma_x \\ \sigma_y \end{pmatrix}, \qquad \Sigma = \begin{pmatrix} \sigma_x^2 & \rho\sigma_x\sigma_y \\ \rho\sigma_x\sigma_y & \sigma_y^2 \end{pmatrix}.$$

In the bivariate case, the first equivalent condition for multivariate normality can be made less restrictive: it is sufficient to verify that countably many distinct linear combinations of ***X*** and ***Y*** are normal in order to conclude that the vector ***[X Y]'*** is bivariate normal.

Gaussian filtering is used to blur images and remove noise and detail. Gaussian filter uses the above Gaussian function which is given in equation 1.

After applying Gauss transform or Wierstrass transform we can compare original false color composite (FCC) image with transformed false color composite (FCC) image.

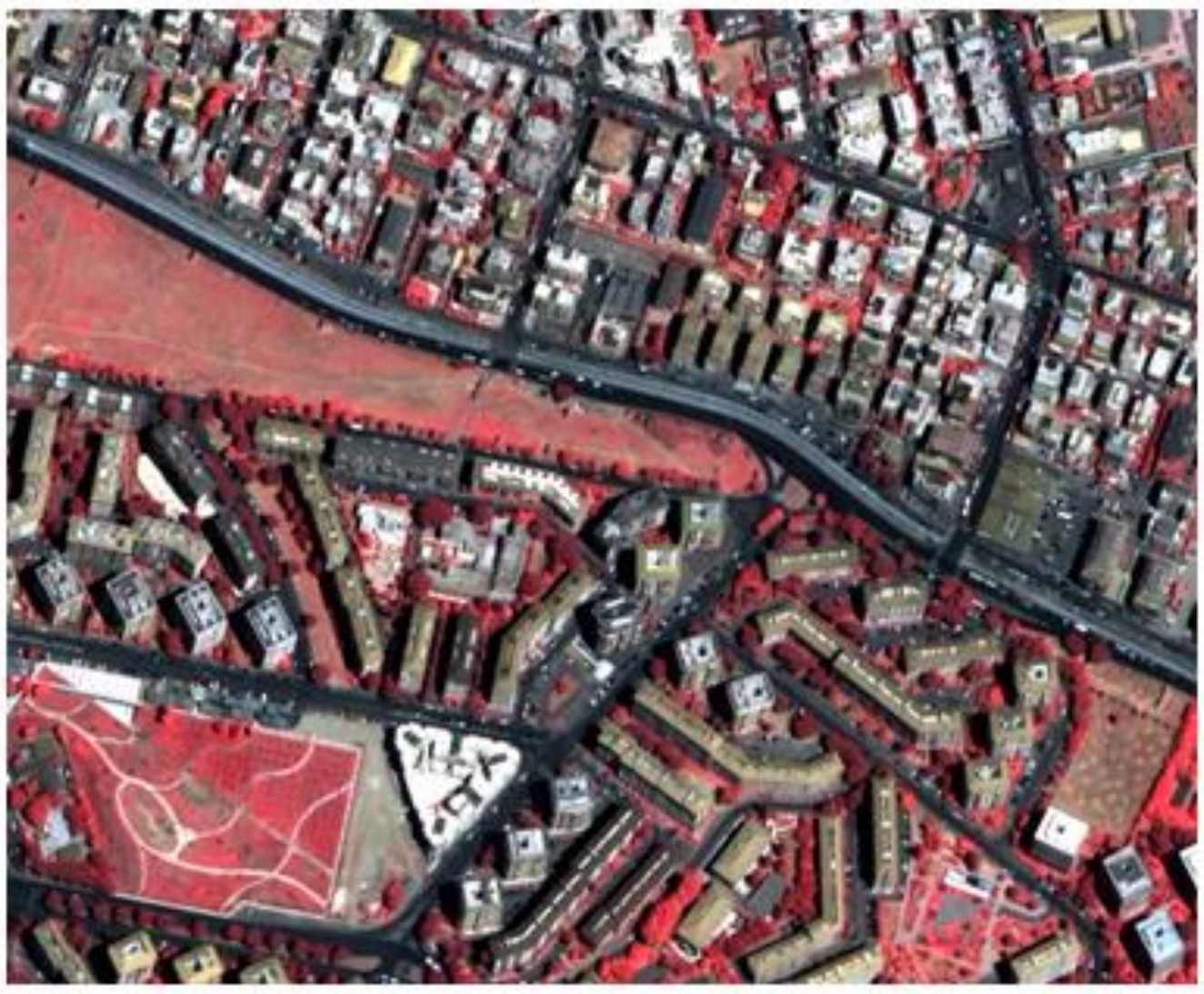

(a) original false color composite (FCC) image

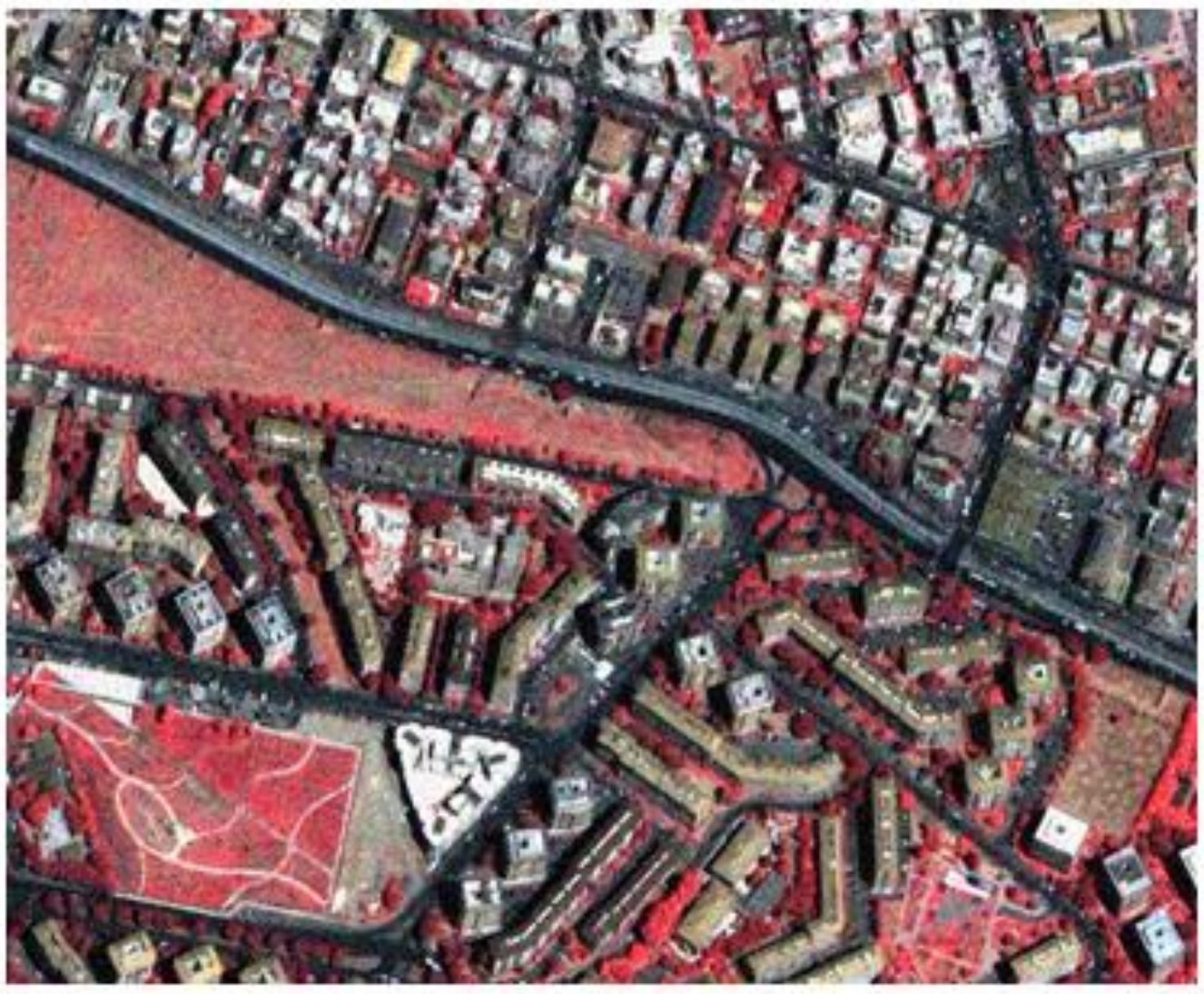

(b) Transformed false color composite (FCC) image

Fig. 3.

In figure 3 we can see the clear difference between original FCC image with transformed FCC image. After applying transformation we can see that the transformed image is more blur than original image.

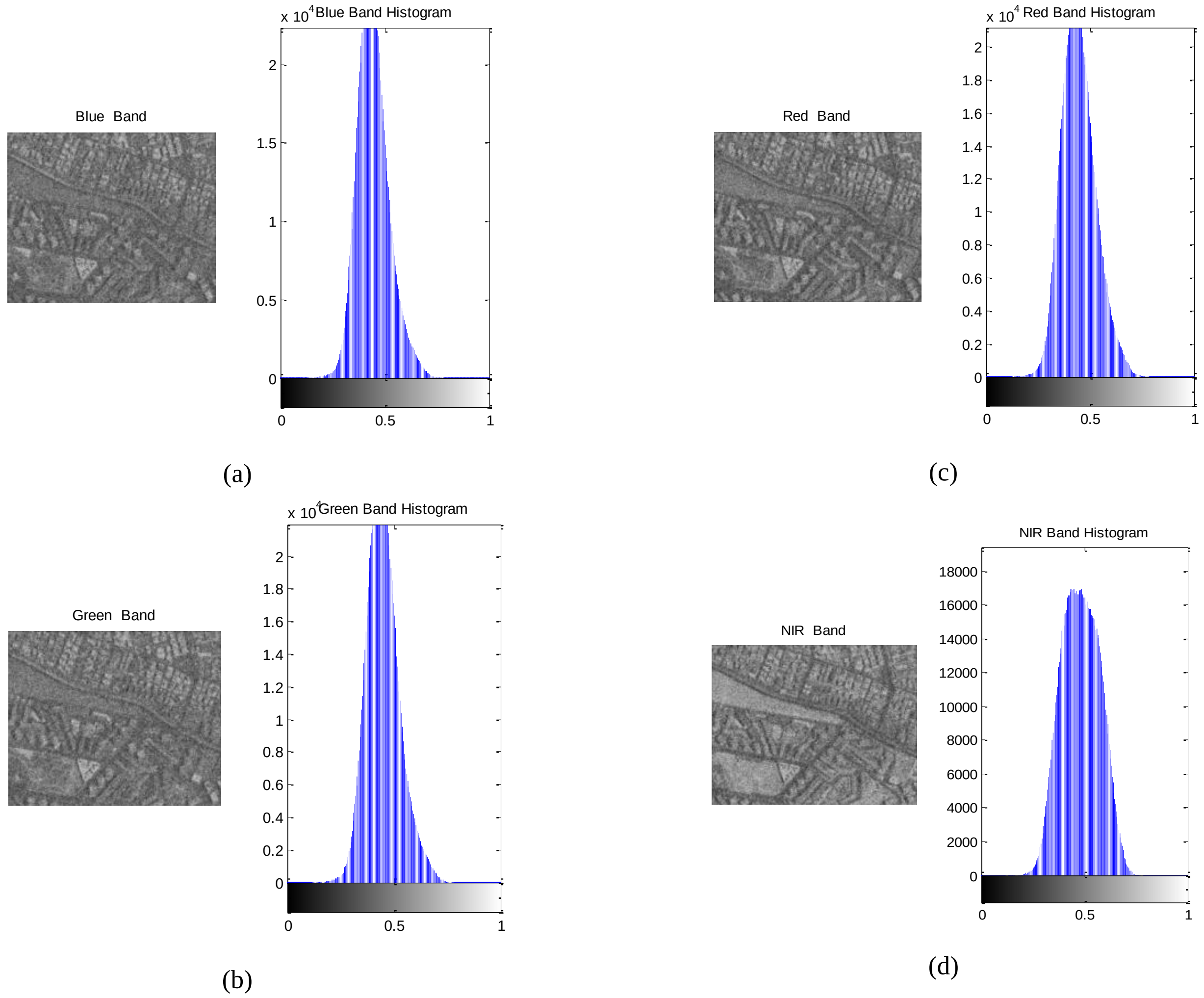

Fig. 4. (a) Histogram of band 1 of transformed image, (b) Histogram of band 2 of transformed image, (c) Histogram of band 3 of transformed image, (d) Histogram of band 4 of transformed image

In Figure 4 if we look at the histograms of different bands of transformed false color composite (FCC) image we can examine that the distribution of each band follow Gaussian distribution. We can see now pixel values of image follow standard Gaussian distribution.

## *C. Dimension Reduction*

Dimension reduction in satellite image processing reducing the number of bands which have same information in nature. Dimension Reduction is used when we have large number of bands.

We have Quickbird image which have 4 bands and we apply PCA to check which band we have to ignore. We apply PCA on transformed FCC image.

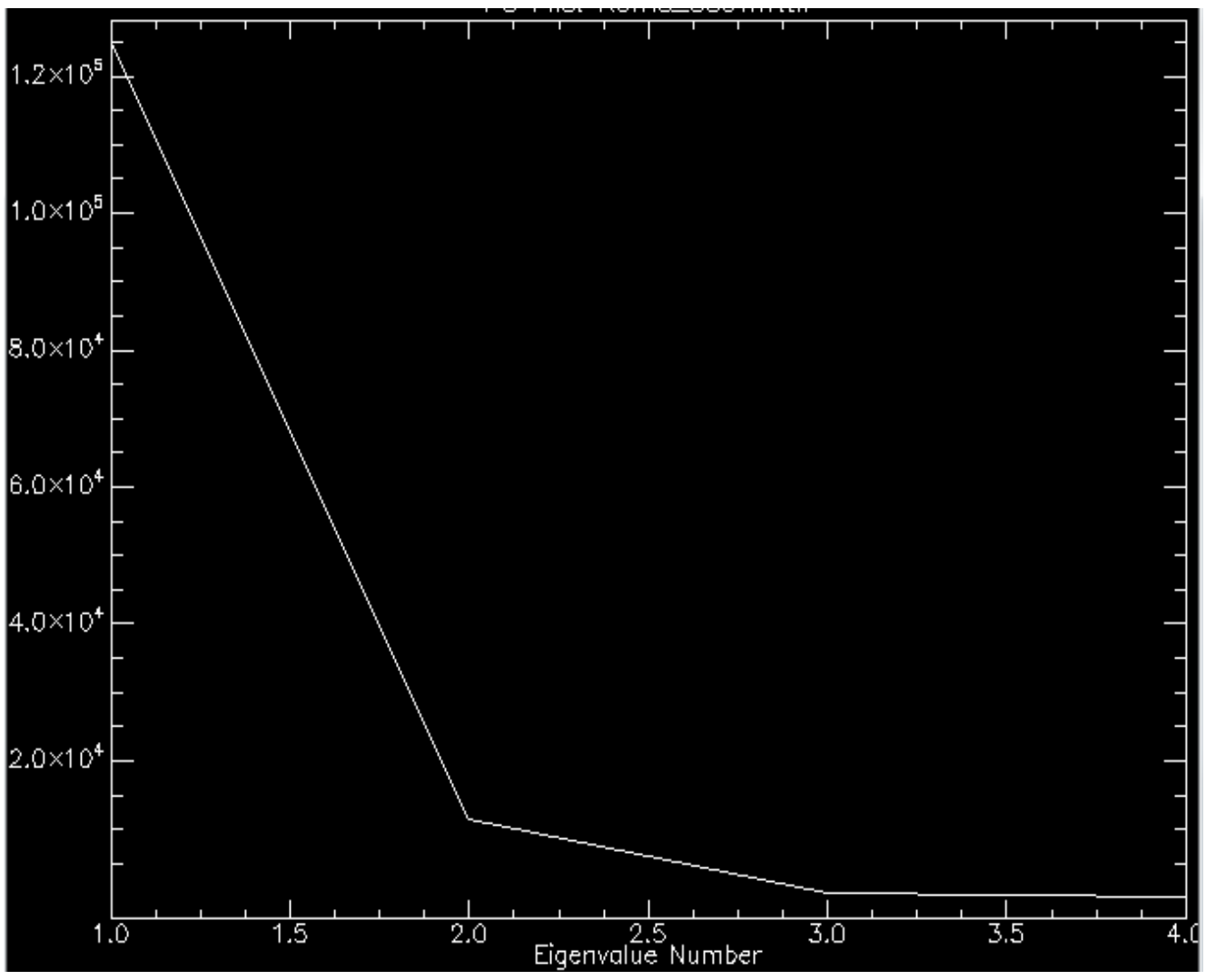

Fig. 4. Principle Components of four bands of transformed image.

In the above figure we can see that band 3(Red) and band 4(NIR) have same information so we have to neglect band 3 or band 4 from our analysis because they have same information in the image for this purpose we will use classification on three bands that are Blue, Green and NIR.

*D. Training Samples*

Training areas were established by choosing one or more polygons for each class. Pixels fall within the training area were taken to be the training pixels for a particular class. In order to select a good training area for a class, the important properties taken into consideration are its uniformity and how well they represent the same class throughout the whole image [4].

We use Stratified random sampling for selecting the training pixels. A stratified random sample is a population sample that requires the population to be divided into smaller groups, called 'strata'. Random samples can be taken from each stratum, or group.

## V. Results

Maximum Likelihood considered to be an most accurate algorithm when we compare it to classical algorithm such as parallelepiped classification algorithm. But as we know that ML classification algorithm assumes that distribution of each band has to be Gaussian. we had seen that our band's distribution was not Gaussian so we make distribution Gaussian by applying Gauss transform or Weirstrass transform. Now we will check what is the effect on accuracy of classification classes when the distribution is Gaussian and when not Gaussian.

*A. Classification Before Weirstrass Transform*

The outcome of ML classification after assigning the classes with suitable colors, is shown in Fig.5 (a): Grass (green), Urban Area (blue), Roads(yellow), Soil (cyan), Trees (red) and Unclassified (Black). The areas in terms of percentage and square meters were also computed; the classes with the largest area is Urban Area. In this case the distribution of our data is not Gaussian and we can clearly see in the Fig.5. (a) that some buildings are wrongly classified as Roads because we are applying ML classification on non normal data.

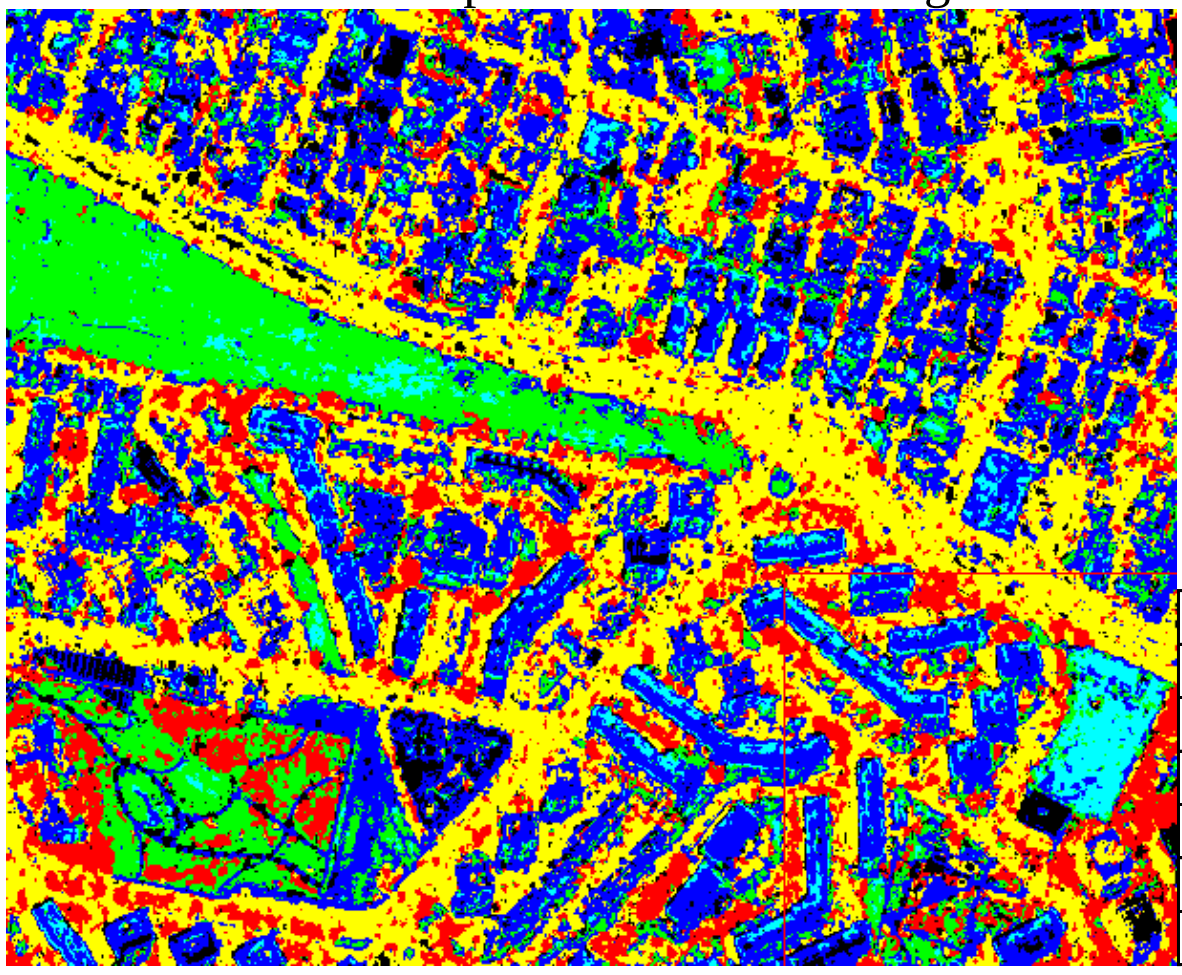

| Class | Color | Area in % | Area in Meters²) | Color |
|---|---|---|---|---|
| Unclassified: | [Black] | 6.44% | (26,783.6400 Meters²) | |
| Trees | [Red] | 14.20% | (59,095.0800 Meters²) | |
| Grass | [Green] | 13.63% | (56,719.0800 Meters²) | |
| Urban Area | [Blue] | 31.45% | (130,868.2800 Meters²) | |
| Roads | [Yellow] | 30.31% | (126,133.9200 Meters²) | |
| Soil | [Cyan] | 3.97% | (16,532.6400 Meters²) | |

(a)

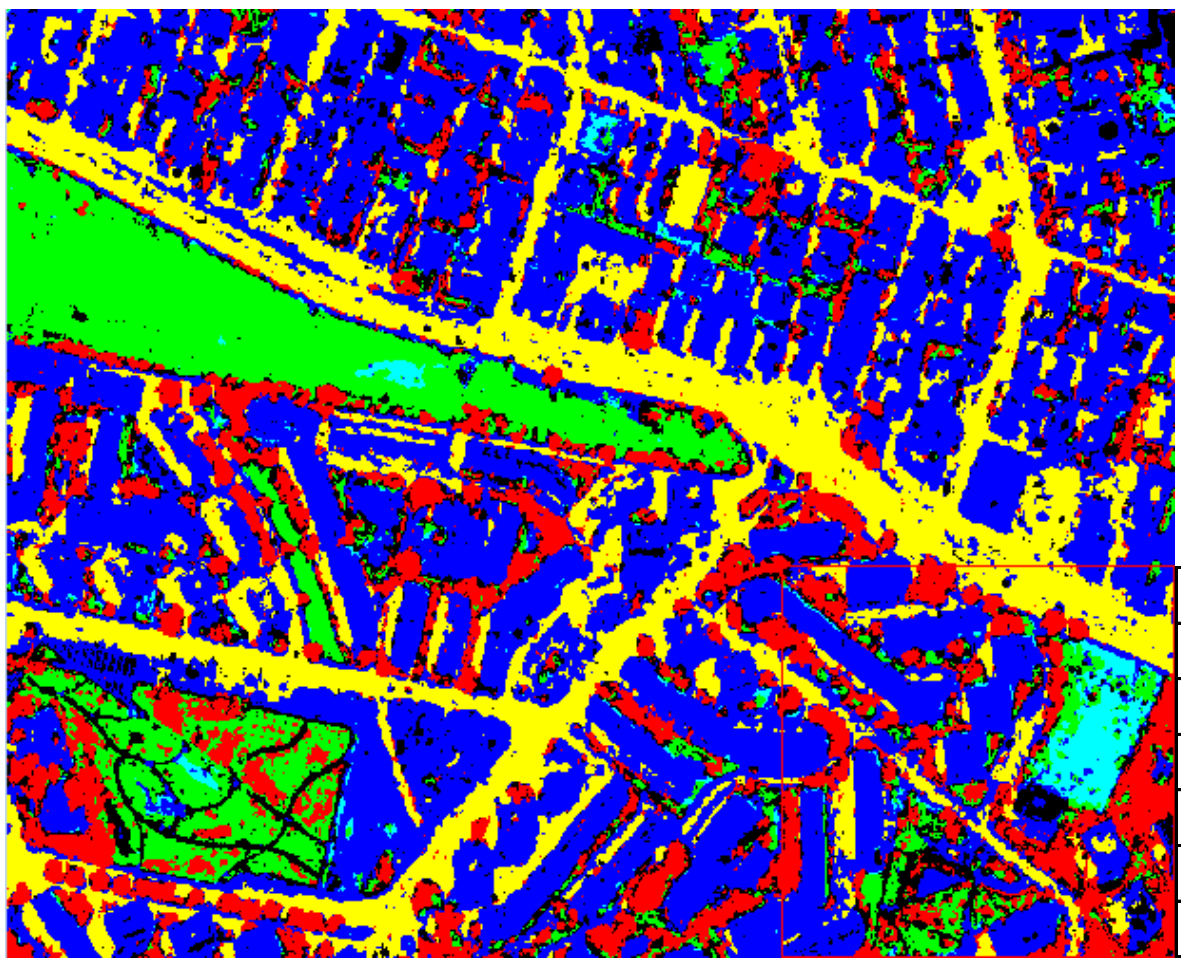

| Class | Color | Area in % | Area in Meters²) | Color |
|---|---|---|---|---|
| Unclassified: | [Black] | 10.30% | (42,851.1600 Meters²) | |
| Trees | [Red] | 13.75% | (57,232.0800 Meters²) | |
| Grass | [Green] | 10.34% | (43,005.2400 Meters²) | |
| Urban Area | [Blue] | 44.79% | (186,373.0800 Meters²) | |
| Roads | [Yellow] | 19.39% | (80,677.4400 Meters²) | |
| Soil | [Cyan] | 1.44% | (5,993.6400 Meters²) | |

(b)

Fig. 5. (a) Maximum Likelihood classification an original image, (b) Maximum Likelihood classification on transformed image

TABLE I. CONFUSION MATRIX OF ORIGINAL CLASSIFIED IMAGE

| Overall Accuracy = (18318/21394) = 85.62% | | | | | | | |
|---|---|---|---|---|---|---|---|
| Kappa Coefficient = 0.8135 | | | | | | | |
| | Ground Truth (Percent) | | | | | | |
| Class | Color | Trees | Grass | Urban Area | Roads | Soil | Color |
| Unclassified | [Black] | 0.4 | 0.05 | 7.84 | 2.33 | 0 | |
| Trees | [Red] | 94.38 | 8.06 | 0.94 | 5.28 | 0 | |
| Grass | [Green] | 1.64 | 88.71 | 2.66 | 0.03 | 3.41 | |
| Urban Area | [Blue] | 0 | 1.06 | 70.71 | 1.19 | 2 | |
| Roads | [Yellow | 3.58 | 0 | 4.61 | 91.17 | 0 | |
| Soil | [Cyan] | 0 | 2.11 | 13.23 | 0 | 94.59 | |
| | Total | 100 | 100 | 100 | 100 | 100 | |

TABLE II. CONFUSION MATRIX OF TRANSFORMED CLASSIFIED IMAGE

| Overall Accuracy = (20156/21394) = 94.2133% | | | | | | | |
|---|---|---|---|---|---|---|---|
| Kappa Coefficient = 0.9236 | | | | | | | |
| | Ground Truth (Percent) | | | | | | |
| Class | Color | Trees | Grass | Urban Area | Roads | Soil | Color |
| Unclassified | [Black] | 0.85 | 0.36 | 4.36 | 0.52 | 0 | |
| Trees | [Red] | 98.51 | 4.72 | 0.09 | 0.08 | 0 | |
| Grass | [Green] | 0.65 | 94.71 | 1.92 | 0 | 0.3 | |
| Urban Area | [Blue] | 0 | 0.2 | 86.5 | 1.33 | 0.52 | |
| Roads | [Yellow] | 0 | 0 | 3.19 | 98.07 | 0 | |
| Soil | [Cyan] | 0 | 0 | 3.94 | 0 | 99.19 | |
| | Total | 100 | 100 | 100 | 100 | 100 | |

Accuracy assessment of the ML classification was determined by means of a confusion matrix (sometimes called error matrix), which compares, on a class-by class basis, the relationship between reference data (ground truth) and the corresponding results of a classification [1].

The results are given in Table 1. The diagonal elements in Table 1 represent the percentage of correctly assigned and are also known as the producer accuracy. Producer accuracy is a measure of the accuracy of a particular classification scheme and shows the percentage of a particular ground class that is correctly classified.

The Kappa coefficient, κ is a second measure of classification accuracy which incorporates the off-diagonal elements as well as the diagonal terms to give a more robust assessment of accuracy than overall accuracy.

The ML classification yielded an overall accuracy of 85.62% and kappa coefficient 0.8135.

### *B. Classification After Weirstrass Transform*

When we apply Gauss transform on data we have already seen that our band's distribution changes from skew symmetric to approximately Gaussian.

When we apply ML classification algorithm on transformed image we have seen that the overall accuracy increases from 85.62% to 94.21% because pixel values of image data follows Gaussian distribution and also the value of Kappa coefficient, κ increases from 0.8135 to 0.9236. In Table 2 we can see the results of classification after applying the Weirstrass Transform or Gauss transform. So when our data follows Gaussian distribution we can say that ML classification produce better.

## VI. CONCLUSIONS

In this study, detail analyses of ML classification for tropical land covers in Rome have been carried out, in which lead to a number of conclusions. ML classifies the classes that exist in the study area with a good agreement when distribution follows Gaussian and when distributions is not Gaussian we can see that overall accuracy decreases almost 9%. ML classified the study area into 5 classes, with accuracy 88% (κ= 0.8425). ML classifies pixels based on known properties of each cover type, but the generated classes may not be statistically separable.

## ACKNOWLEDGMENT

The authors would like to thanks the Dr Waqas A. Qazi for his valuable input and ideas, Muhammad Qasim for helping with the implementation of Weirstrass transform.

**Muhammad Shoaib** is Master student of RS and GIS in the department of Space Science in Institute of Space Technology Islamabad, Pakistan.

**Zaka Ur Rehman** is Master student of RS and GIS in the department of Space Science in Institute of Space Technology Islamabad, Pakistan.

**Muhammad Qasim** received the Masters degree in Mathematics from Comsats Institute of information technology Abbottabad, Pakistan.